\documentclass[aps,prd,twocolumn,showpacs]{revtex4}
\def\tabletextsize{} 

\def\be{\beta}
\def\ga{\gamma}
\def\de{\delta}

\def\ve{\varepsilon}

\def\et{\eta}
\def\th{\theta}

\def\ka{\kappa}
\def\la{\lambda}

\def\ch{\chi}

\def\om{\omega}

\def\De{\Delta}

\def\Om{\Omega}

\def\fr#1#2{{{#1} \over {#2}}}
\def\frac#1#2{{\textstyle{{#1}\over {#2}}}}
\def\half{{\textstyle{1\over 2}}}
\def\lsim{\mathrel{\rlap{\lower4pt\hbox{\hskip1pt$\sim$}}
    \raise1pt\hbox{$<$}}}
\def\gsim{\mathrel{\rlap{\lower4pt\hbox{\hskip1pt$\sim$}}
    \raise1pt\hbox{$>$}}}
\def\sqr#1#2{{\vcenter{\vbox{\hrule height.#2pt
         \hbox{\vrule width.#2pt height#1pt \kern#1pt
         \vrule width.#2pt}
         \hrule height.#2pt}}}}

\def\pr#1{{#1}^\prime}

\def\abs#1{\left|{#1}\right|}

\def\tb{\tilde{b}} 
\def\td{\tilde{d}} 
\def\tg{\tilde{g}} 
\def\tc{\tilde{c}} 
\def\tH{\tilde{H}} 

\def\tmf{ \widetilde{m}_F }
\def\hmf{ \widehat{m}_F }

\def\dnbeam{\de\bar{\nu}_{\rm beam}}
\def\dnlab{\de\nu_{\rm lab}}

\def\nop{-}

\def\etal{{\it et al.}}

\def\bet{$\beta$} 
\def\bes{$\beta^2$} 
\def\beo{$\beta\beta_\oplus$} 
\def\bto{$\beta^2\beta_\oplus$} 
 
\def\btl{$\beta\beta_L$} 
 
\def\nop{$\cdot$} 

\def\codt{\cos{\omega_\oplus T_\oplus}}
\def\sodt{\sin{\omega_\oplus T_\oplus}}
\def\ctodt{\cos{2\omega_\oplus T_\oplus}}
\def\stodt{\sin{2\omega_\oplus T_\oplus}} 
 
\def\cx{c_\xi} 
\def\sx{s_\xi} 
\def\ce{c_\eta} 
\def\se{s_\eta} 
\def\cto{c_{\Omega T}} 
\def\sto{s_{\Omega T}} 
\def\ctx{c_{2\xi}} 
\def\stx{s_{2\xi}}

 \def\bx{\tilde{b}_{X}} 
 \def\by{\tilde{b}_{Y}} 
 \def\bz{\tilde{b}_{Z}} 
 \def\dx{\tilde{d}_{X}} 
 \def\dy{\tilde{d}_{Z}} 
 \def\dz{\tilde{d}_{Z}} 
\def\gtx{\tilde{g}_{TX}} 
\def\gty{\tilde{g}_{TY}} 
\def\gtz{\tilde{g}_{TZ}} 
 \def\gm{\tilde{g}_{-}} 
\def\gdx{\tilde{g}_{DX}} 
\def\gdy{\tilde{g}_{DY}} 
\def\gdz{\tilde{g}_{DZ}} 
 \def\gq{\tilde{g}_{Q}} 
\def\hxt{\tilde{H}_{XT}} 
\def\hyt{\tilde{H}_{YT}} 
\def\hzt{\tilde{H}_{ZT}} 
\def\dxy{\tilde{d}_{XY}} 
\def\dyz{\tilde{d}_{YZ}} 
\def\dzx{\tilde{d}_{ZX}}

 \def\dm{\tilde{d}_{-}} 
 \def\bt{\tilde{b}_{T}} 
 \def\gt{\tilde{g}_{T}} 
 \def\gc{\tilde{g}_{c}} 
\def\gxy{\tilde{g}_{XY}} 
\def\gxz{\tilde{g}_{XZ}} 
\def\gyx{\tilde{g}_{YX}} 
\def\gyz{\tilde{g}_{YZ}} 
\def\gzx{\tilde{g}_{ZX}} 
\def\gzy{\tilde{g}_{ZY}} 

\def\tcq{\tilde{c}_{Q}} 
\def\tcx{\tilde{c}_{X}} 
\def\tcy{\tilde{c}_{Y}} 
\def\tcz{\tilde{c}_{Z}} 
\def\tcm{\tilde{c}_{-}} 
\def\tctx{\tilde{c}_{TX}} 
\def\tcty{\tilde{c}_{TY}} 
\def\tctz{\tilde{c}_{TZ}} 

\def\citelabel#1{ \label{#1} } 
\def\labeleeq#1{ \label{#1} \end{equation} } 
\def\labeleea#1{ \label{#1} \end{eqnarray} } 
\def\citebibitem#1{ \bibitem{#1} } 

\newcommand{\beq}{\begin{equation}}
\newcommand{\eeq}{\end{equation}}
\newcommand{\bea}{\begin{eqnarray}}
\newcommand{\eea}{\end{eqnarray}}

\newcommand{\Eq}[1]{Eq.\ (\ref{#1})}

\begin{document}

\preprint{Berry College HEP-002}

\title{Probing Lorentz violation with Doppler-shift experiments}

\author{Charles D.\ Lane}
 \email{clane@berry.edu}
\affiliation{Physics Department, Berry College, Mount Berry, GA 30149-5004} 

\date{\today}

\begin{abstract} 
This work analyzes Doppler-effect experiments in terms of a general framework for 
violations of Lorentz symmetry: 
the Standard-Model Extension. 
These experiments are found to be sensitive to 
heretofore unprobed combinations of Lorentz-violation coefficients 
associated with protons and electrons. 
New bounds at the level of $10^{-11}$ and $10^{-8}$ for proton coefficients 
and $10^{-2}$ for electron coefficients 
emerge from a recent experiment. 
\end{abstract} 

\pacs{03.30.+p,06.30.Ft,11.30.Cp,11.30.Er}% PACS, the Physics and Astronomy
                             % Classification Scheme.
%\keywords{Suggested keywords}%Use showkeys class option if keyword
                              %display desired
\maketitle

%=================================================================
%=================================================================

%======================================================================================
\section{Introduction}
%======================================================================================

The Doppler effect of light is one of the classic tests of special relativity \cite{IvesStilwell}, 
particularly symmetry under Lorentz transformations. 
Modern experiments agree with relativistic predictions 
to a very high degree of precision \cite{doppler1}. 
However, the possibility of Lorentz symmetry violation 
due to Planck-scale physics \cite{cpt04,mattingly} 
makes continuing study of the relativistic Doppler shift an important endeavor. 

This work uses the nongravitational sector of the Standard-Model Extension (SME), 
a general framework for violations of particle Lorentz symmetry, 
to analyze Doppler-effect experiments. 
Most previous analyses have used a limited test model \cite{RMS} 
that can describe only a single type of matter aside from the photon, 
and that is likely a special case of the SME \cite{SMEelectrodynamics1}. 
Analysis of the most recent Doppler experiment \cite{doppler1} 
has been performed \cite{tobar} within the context of the photon sector of the SME, 
describing sensitivity to a combination of photon parameters; 
the photon sector is neglected in most of the current paper. 
Analysis within the context of the fermion sector of the SME 
has yet to be performed. 
Herein is presented such an analysis, yielding new bounds on a set of 
proton- and electron-associated coefficients for Lorentz violation. 
This work also indicates modifications to current experiments 
that could yield sensitivity to a much broader class of coefficients. 

Theoretical analyses of the SME include 
its possible origin in spontaneously broken Lorentz symmetry \cite{SMEssb}, 
its definition \cite{SMEdefn}, 
its spin-statistics relation and microcausal structure \cite{SMEmicrocausal}, 
its renormalizability \cite{SMErenorm}, 
its extension to gravitational physics \cite{SMEgravity}, 
and its hamiltonian for free Dirac fermions \cite{SMEhamiltonian1,SMEhamiltonian2}. 
Phenomenological studies of the SME include analyses of 
the electron sector \cite{SMEelectrons1,SMEelectrons2}, 
atomic clocks \cite{SMEclocks1,SMEclocks2,SMEclocks3}, 
electrodynamics \cite{SMEelectrons2,SMEelectrodynamics1,SMEelectrodynamics2,SMEelectrodynamics3}, 
muon behavior \cite{SMEmuons}, 
neutral mesons \cite{SMEmesons}, 
neutrinos \cite{SMEneutrinos}, 
the Higgs field \cite{SMEhiggs}, 
and more. 
Most of these phenomenological studies have been applied to existing experiments, 
including those with 
the electron sector \cite{EXPTelectrons}, 
atomic clocks \cite{EXPTclocks}, 
electrodynamics \cite{EXPTelectrodynamics}, 
muon behavior \cite{EXPTmuons}, 
neutral mesons \cite{EXPTmesons}, 
and neutrinos \cite{EXPTneutrinos}. 

This paper is organized as follows. 
Section \ref{Basics} collects many small discussions: 
the ingredients of a Doppler-effect experiment, 
qualitative reasons for expecting such experiments to be sensitive to Lorentz violation, 
definition of the key experimental quantity of interest, 
and relevant coordinate systems for detailed analysis. 
Section \ref{SMEPredictions} describes the signals that would appear in Doppler experiments 
were Lorentz symmetry to be violated 
in accordance with the SME. 
The general results are applied to the completed Heidelberg experiment 
in Sec.\ \ref{Heidelberg}. 
Section \ref{FutureExperiments} discusses a few ways that future experiments 
could yield sensitivity to a broader set of Lorentz-violation coefficients. 
A summary appears in Sec.\ \ref{Summary}, 
and an Appendix contains some calculational details. 

%======================================================================================
\section{Basics}
\citelabel{Basics} 
%======================================================================================

Observer Lorentz symmetry is maintained in the 
SME, 
and hence conventional relationships hold 
between measurements made in different inertial frames 
%of a single instance of any given physical quantity. 
of the same physical quantity. 
However, 
due to violation of particle Lorentz symmetry, 
analysis of experimental conditions has an extra subtlety. 
To wit: Since measurements made in identical experimental setups 
moving with respect to each other 
do {\itshape not} measure the same physical quantity, 
but rather different instances of similar physical quantities, 
they may not obey conventional relationships. 
For example, the rest-frame frequency associated with an atomic transition 
depends on the particular frame of the atom, 
in contrast to the conventional Lorentz-symmetric case. 

\subsection{Experimental basics} 
Doppler-shift experiments involve three main ingredients. 
The first ingredient consists of two groups of atoms, 
one group at rest in the lab frame 
and the other forming a beam with velocity $\vec{\be}$ in the lab frame. 
The same transition should be studied in each group. 
A pair of coordinate frames, 
the lab frame ${\cal S}$ and the moving atoms' rest frame $\bar{\cal S}$, 
form the second ingredient. 
The third ingredient is a set of 
three transition-frequency measurements: 
\begin{enumerate} 
\item[] $\nu_{\rm lab}$= frequency associated with the atoms at rest in the lab frame, 
 measured in the lab frame (and hence in their own rest frame), 
\item[] $\bar{\nu}_{\rm beam}$= frequency associated with the beam atoms, 
 measured in the beam frame (and hence in their own rest frame), and  
\item[] $\nu_{\rm beam}$= frequency associated with the beam atoms, 
 measured in the lab frame ({\itshape not} their own rest frame). 
\end{enumerate} 

%\subsection{Qualitative reasoning} 
General considerations of Lorentz symmetry give qualitative insight 
into relationships within each of two pairs of the measurements. 
First, consider $\nu_{\rm beam}$ and $\bar{\nu}_{\rm beam}$. 
These are measurements of the same transition in the same atoms,
but measured in different frames, 
so they are related by an {\itshape observer} Lorentz transformation. 
Since the SME preserves observer Lorentz symmetry, 
$\nu_{\rm beam}$ and $\bar{\nu}_{\rm beam}$ obey their conventional Doppler relationship 
in this theory. 

Next, consider $\nu_{\rm lab}$ and $\bar{\nu}_{\rm beam}$. 
These are not measurements of the same quantity. 
However, since they are measurements of the same type of thing 
(a particular transition in a particular type of atom), 
they are related by a {\itshape particle} Lorentz transformation \cite{FootnoteObsPart}. 
Since the SME breaks particle Lorentz symmetry, 
they do not obey their conventional relationship, namely, equality. 
Doppler-effect experiments effectively probe the ratio of these two frequency measurements, 
bounding the amount by which it differs from unity. 

In practice, $\nu_{\rm beam}$ is not a single frequency measurement. 
Two different beam-atom transition frequencies are measured in the lab frame: 
$\nu_p$, the frequency of light 
measured by a laser running parallel to the beam in the lab frame, 
and $\nu_a$, the frequency of light 
measured by a laser running antiparallel to the beam in the lab frame. 
In conventional physics, a nice identity relates these to the beam transition frequency 
as measured in the beam frame: $\nu_a \nu_p = \bar{\nu}_{\rm beam}^2$. 
This relationship is based only on observer Lorentz transformations, 
and therefore holds in both conventional physics and the SME. 
Because of this, 
the analysis may be performed as if the frequencies $\bar{\nu}_{\rm beam}$ and $\nu_{\rm lab}$ 
are directly measured, 
even though the actual experiments deal with frequencies $\nu_a$, $\nu_p$, and $\nu_{\rm lab}$. 

The key experimental quantity is the ratio $\nu_a\nu_p/\nu_{\rm lab}^2$. 
Due to the identity mentioned above, this can be expressed as 
the square of the ratio of two transition frequencies 
as measured in their own rest frames, 
$\bar{\nu}_{\rm beam}$ and $\nu_{\rm lab}$. 
As discussed above, these frequencies 
are equal in conventional physics, 
but may be different whenever particle Lorentz symmetry is violated. 
Thus, in conventional physics, the experimental quantity of interest is exactly unity. 
When Lorentz symmetry is violated, however, 
it may differ slightly from one. 
Moreover, for experiments conducted in a frame that is not strictly inertial 
(such as those based on Earth's surface), 
the key ratio may vary with time. 

Since particle Lorentz symmetry is at least approximately valid, 
the experimental quantity of interest can conveniently be written 
as a perturbation around one: 
\beq 
\fr{\nu_a\nu_p}{\nu_{\rm lab}^2} 
 = \fr{\bar{\nu}_{\rm beam}^2}{\nu_{\rm lab}^2} 
 = 1 + \ve 
\quad , 
\labeleeq{exptquant1} 
where $\ve\ll 1$. 
Section \ref{SMEPredictions} of this paper describes the size and time dependence of $\ve$ 
under general experimental scenarios. 

The most stringent bound on $\ve$ to date results from a recent Doppler-effect experiment 
at the heavy-ion storage ring TSR in Heidelberg \cite{doppler1}. 
This experiment limits the magnitude of $\ve$ to 
\beq 
\abs{\ve} \lsim 2\times 10^{-9} 
\quad . 
\labeleeq{bound1} 

\subsection{Coordinate frames.} 
Three coordinate frames are necessary for this work. 
The first is a nonrotating system $(T,X,Y,Z)$ 
centered on the Sun. 
This frame is approximately inertial over the course of 
any realistic Doppler-shift experiment. 
It is defined in detail in Section III of Ref.\ \cite{SMEclocks3}. 

The second system $(t,x,y,z)$ is fixed in the laboratory 
on Earth's surface. 
In it, the $x$ axis points south, 
the $y$ axis points east, 
and the $z$ axis points vertically upwards. 
Useful angles in this frame include the angle $\et\approx 23.4^\circ$ between Earth's 
equatorial and orbital planes 
and the colatitude $\ch$ of the laboratory. 
The angular frequency of Earth's revolution about the Sun is denoted $\Om_\oplus$, 
while the rotational frequency of the Earth is denoted $\om_\oplus$. 
The current work uses $\be_\oplus\approx 10^{-4}$ to denote the average speed of Earth with respect to the Sun, 
and $\be_L\approx (1.5\times 10^{-6})\sin{\ch}$ to denote 
the average speed of the lab with respect to Earth's rotational axis. 
These first two systems and transformations between them 
are defined in detail in Appendix C of Ref.\ \cite{SMEelectrodynamics1}. 

The third frame $(\bar t,\bar x,\bar y,\bar z)$ is the rest frame of the beam atoms. 
These axes are defined to move with speed $\be$ along the lab-frame $y$-axis. 
The $\bar y$ axis is chosen to be parallel to the $y$ axis. 
The $\bar z$ axis points along the quantization axis of the beam atoms, 
which is assumed to be rotated around the $y$ axis by angle $\th$ 
away from the $z$ axis. 
Thus, for an experiment conducted at a colatitude $\ch$, 
the angle between Earth's rotation axis and the quantization axis 
is given by $\xi=\ch+\th$. 
The current work doesn't include 
generalization to beams moving with arbitrary orientation: 
such generalization is straightforward 
and has little effect on the analysis aside from computational messiness. 
The $\bar x$ axis is defined so as to complete the right-handed system. 

This work keeps only terms up to second order in $\be$ for two reasons: 
First, for all existing Doppler-effect experiments to date, $\be\ll 1$. 
In the most recent experiment \cite{doppler1}, for example, $\be\approx 0.064$. 
Second, the types of Lorentz violation to which Doppler-effect experiments are sensitive 
are unaffected by inclusion of higher-order terms. 
Moreover, at most first-order terms in $\be_\oplus$ and $\be_L$ are kept 
since each is much smaller than $\be$.

%======================================================================================
\section{SME Predictions} 
\citelabel{SMEPredictions} 
%======================================================================================
In this section, 
the behavior of the key experimental quantity is described 
as predicted by the SME. 
First, the shift to an atom's transition frequency is considered 
in an approximately inertial frame instantaneously at rest with the atom. 
Next, the frequencies of atoms moving at constant velocity with respect to each other are compared. 
The section concludes with the consequences of the motion of the atoms 
as they rotate with Earth and ride Earth around the Sun. 

%-------------------------------------------------------------------
\subsection{Approximately inertial frame} 
\citelabel{InerFrame} 
%-------------------------------------------------------------------
In the SME, 
new interactions cause shifts in atomic energy levels 
with respect to conventional values \cite{SMEclocks1}. 
The energy-level shift $\de E$ to an atomic state with total-angular-momentum quantum number $F$ 
and projection number $m_F$ takes the form 
$\de E(F,m_F) = \hmf E_d(F) + \tmf E_q(F)$. 
In this expression, $\hmf$ and $\tmf$ are ratios of Clebsch-Gordan coefficients 
while $E_d(F)$ and $E_q(F)$ are $m_F$-independent dipole- and quadrupole-type energy-level shifts. 
%In this expression, $\hmf:=m_F/F$ and $\tmf:=[3m_F^2-F(F+1)]/[3F^2-F(F+1)]$ 
%are ratios of Clebsch-Gordan coefficients 
%while $E_d(F)=\sum_w(\be_w\tb_3^w+\de_w\tb_3^w+\ka_w\ga_d^w)$, 
%$E_q(F)=\sum_w(\ga_w\tc_q^w+\la_w\tg_q^w)$ are $m_F$-independent dipole- and quadrupole-type energy-level shifts. 
%The sums in the energy-level shifts are over particle species proton, neutron, and electron. 

In turn, this generally induces transition-frequency shifts of the form 
\begin{eqnarray} 
\nu &=& \nu_{\rm SM} + \de\nu \quad, \nonumber \\ 
\de\nu &=& \fr{1}{2\pi} \sum_w \Big[ 
  (\hmf\be_w-\pr{\hmf}\pr{\be}_w) \tb_3^w \nonumber \\ 
&& \qquad\quad +(\hmf\de_w-\pr{\hmf}\pr{\de}_w) \td_3^w \nonumber \\ 
&& \qquad\quad +(\hmf\ka_w-\pr{\hmf}\pr{\ka}_w) \tg_d^w \nonumber \\ 
&& \qquad\quad +(\tmf\ga_w-\pr{\tmf}\pr{\ga}_w) \tc_q^w \nonumber \\ 
&& \qquad\quad +(\tmf\la_w-\pr{\tmf}\pr{\la}_w) \tg_q^w  \Big] 
\quad , 
\labeleea{Denu} 
where $\nu$ denotes any transition frequency measured in the atom's own rest frame, 
assuming that the tensor components are evaluated in that frame; 
$\nu_{\rm SM}$ is the conventional (i.e., according-to-the-standard-model) frequency; 
and $\de\nu$ is the shift induced by the SME. 
The sum is over the particle types proton, neutron, and electron. 
The quantities $\be_w$, \ldots, $\la_w$ that describe $m_F$-independent 
properties of the atoms are described in the paragraphs surrounding 
Eqs.\ (10) through (12) of Ref.\ \cite{SMEclocks1}. 
Finally, $\tb_3^w$, \ldots, $\tg_q^w$ are the specific combinations of 
Lorentz-violation tensor components 
that appear in studies of atomic systems; 
explicit definitions of them appear as Eq.\ (9) of Ref.\ \cite{SMEclocks1}. 

\subsection{Comparison of lab and beam atoms} 
Expression (\ref{Denu}) holds for both $\nu_{\rm lab}$ and $\bar{\nu}_{\rm beam}$, 
with the tensor components evaluated in the lab and beam frames, respectively. 
Since tensor components generally depend on the frame in which they're evaluated, 
it usually pertains that $\tb_{\bar{3}}\ne \tb_{3}$, $\tc_{\bar{q}}\ne \tc_{q}$, etc. 
Thus, since they depend on tensor components evaluated in different frames, 
$\dnbeam\ne \dnlab$ and 
hence $\bar{\nu}_{\rm beam}\ne \nu_{\rm lab}$. 

Since Doppler-effect experiments involve the same transitions in the same atomic species 
(albeit measured in different frames), 
the values of $\hmf$, $\tmf$, and $\be_w$, \ldots, $\la_w$ 
that appear in $\dnbeam$ and $\dnlab$ are identical. 
Thus, 
the experimental quantity of interest becomes 
\bea 
\fr{\bar\nu_{\rm beam}^2}{\nu_{\rm lab}^2} 
 &=& 1+\fr{2}{\nu_{\rm SM}} \left(\de\bar{\nu}_{\rm beam}-\de\nu_{\rm lab} \right) \nonumber \\ 
&=& 1+ \fr{1}{\pi\nu_{\rm SM}} \sum_w 
 \Big[ 
   (\hmf\be_w-\pr{\hmf}\pr{\be}_w) (\tb_{\bar 3}-\tb_3) \nonumber \\ 
&& \quad\qquad +(\hmf\de_w-\pr{\hmf}\pr{\de}_w) (\td_{\bar 3}-\td_3) \nonumber \\ 
&& \quad\qquad +(\hmf\ka_w-\pr{\hmf}\pr{\ka}_w) (\tg_{\bar d}-\tg_d) \nonumber \\ 
&& \quad\qquad +(\tmf\ga_w-\pr{\tmf}\pr{\ga}_w) (\tc_{\bar q}-\tc_q) \nonumber \\ 
&& \quad\qquad +(\tmf\la_w-\pr{\tmf}\pr{\la}_w) (\tg_{\bar q}-\tg_q) 
  \Big] 
\quad . 
\labeleea{exptquant2} 
This quantity varies with time as Earth rotates. 
Its exact time dependence is calculated in Section \ref{NoninerFrame}. 

Some of the coefficients that appear in \Eq{exptquant2} are given in the lab frame, 
while others are given in comoving beam coordinates. 
Conventional Lorentz boosts may be used to express the beam components 
in terms of the lab-frame components. 
To second order in the beam speed $\be$ 
and expressed entirely in lab-frame coordinates, 
the combinations of coefficients in \Eq{exptquant2} are 
\bea 
\tb_{\bar 3}-\tb_3 
 &=& \be\left[ -md_{32}+mg_{122}+mg_{100}-H_{10} \right] \nonumber \\ 
&& + \be^2\left[ -\frac{1}{2}md_{30}+mg_{120}+mg_{102}-\frac{1}{2}H_{12} \right] \quad , \nonumber \\ 
\td_{\bar 3}-\td_3 
 &=& \be\left[ md_{23}+\frac{1}{2}md_{32}-\frac{1}{2}H_{10} \right] \nonumber \\ 
&& + \be^2\left[ \frac{1}{4}md_{30}+\frac{1}{2}md_{03}-\frac{1}{4}H_{12} \right] \quad , \nonumber \\ 
\tg_{\bar d}-\tg_d 
 &=& \be\left[ 2mg_{122}+2mg_{100} \right] \nonumber \\ 
&& + \be^2\left[ 2mg_{120}+2mg_{102} \right] \quad , \nonumber \\ 
\tc_{\bar q}-\tc_q 
 &=& \be\left[ mc_{20}+mc_{02} \right] \nonumber \\ 
&& + \be^2\left[ mc_{11}+2mc_{22}+mc_{33} \right] \quad , \quad {\rm and} \nonumber \\ 
\tg_{\bar q}-\tg_q 
 &=& \be\left[ -mg_{211}+2mg_{233}+mg_{200} \right] \nonumber \\ 
&& + \be^2\left[ \frac{1}{2}mg_{101}+\frac{1}{2}mg_{202}-mg_{303} \right] 
\quad . 
\labeleea{DeltaTildes1} 

%-------------------------------------------------------------------
\subsection{Noninertial frame}
\citelabel{NoninerFrame} 
%-------------------------------------------------------------------
As the coefficients that appear in \Eq{DeltaTildes1} are tensor components 
in a frame attached to Earth's surface, 
they vary as Earth rotates and accelerates through the nonrotating Sun frame. 
Thus, it is best to re-express the key experimental quantity 
in terms of tensor components in the Sun frame, 
thereby explicitly displaying its time dependence. 
In this section, 
$\tc_{\bar q}-\tc_q$ is first discussed in detail, 
then relevant information for all coefficients is summarized. 

In terms of nonrotating-frame components, 
\begin{widetext} 
\bea 
%=====================================================================================
% Begin \tc_q 
%=====================================================================================
\tc_{\bar q}-\tc_q &=& 
% Begin cos(\om T_s) dependence====================================
\codt \Big\{ \be\Big[ 
 \tcty 
 \Big]
 +\be\be_\oplus \Big[ 
 \ce\cto\tcm -\frac{1}{3}\ce\cto\tcq -\se\cto\tcx +\sto\tcz 
 \Big] 
 +\be^2\be_L \Big[ 
 2\tcty 
 \Big] 
 \Big\} 
 \nonumber \\ 
% Begin sin(\om T_s) dependence====================================
&+& \sodt \Big\{ \be\Big[ 
 -\tctx 
 \Big]
 +\be\be_\oplus \Big[ 
 -\sto\tcm -\frac{1}{3}\sto\tcq +\se\cto\tcy +\ce\cto\tcz 
 \Big] 
 +\be^2\be_L \Big[ 
 -2\tctx 
 \Big] 
 \Big\} 
 \nonumber \\ 
% Begin cos(2\om T_s) dependence====================================
&+& \ctodt \Big\{ 
 \be^2 \Big[ 
 -\half\tcm 
 \Big] 
 +\be\be_L \Big[ 
 -\tcm 
 \Big] 
 +\be^2\be_\oplus \Big[ 
 -\half\sto\tctx -\half\ce\cto\tcty 
 \Big] 
 \Big\} 
 \nonumber \\ 
% Begin sin(2\om T_s) dependence====================================
&+& \stodt \Big\{ 
 \be^2 \Big[ 
 -\half\tcz 
 \Big] 
 +\be\be_L \Big[ 
 -\tcz 
 \Big] 
 +\be^2\be_\oplus \Big[ 
 \half\ce\cto\tctx -\half\sto\tcty 
 \Big] 
 \Big\} 
 \nonumber \\ 
% Begin T independence====================================
&+& \Big\{ 
 \be^2 \Big[ 
 \frac{1}{6}\tcq 
 \Big] 
 +\be\be_L \Big[ 
 \frac{1}{3}\tcq 
 \Big] 
 +\be^2\be_\oplus \Big[ 
 \frac{3}{2}\sto\tctx -\frac{3}{2}\ce\cto\tcty -\se\cto\tctz 
 \Big] 
 \Big\} 
\quad . 
\labeleea{Decq} 

\end{widetext} 
In this and later equations, the abbreviations $c_x:=\cos{x}$ and $s_x:=\sin{x}$ are used. 
Further, the combination $\Om_\oplus T$ is shortened to $\Om T$. 
The individual terms in this expression have been sorted into five pieces 
based on their variation at Earth's relatively rapid rotational frequency $\om_\oplus$: 
\bea 
\tc_{\bar q}-\tc_q &=& C_{1}\codt +S_{1}\sodt \nonumber \\ 
 &+& C_{2}\ctodt + S_{2}\stodt +K_0 
\quad , 
\labeleea{DeltaTildes2} 
where the coefficients of these sinusoids and the piece $K_0$ 
include both terms that vary at the relatively slow 
frequency $\Om_\oplus$ of Earth's motion around the Sun 
and terms that are truly constant. 
All combinations of Sun-frame components $\tc_Q$, $\tc_{TX}$, etc. 
in this and later expressions 
are defined in Appendix B of \cite{SMEclocks3}. 

The relevant content of \Eq{Decq} 
deals with the $\om_\oplus$ time dependence, 
the set of nonrotating-frame parameters on which it depends, 
and the suppression of each of these parameters by products of various speeds. 
Exact dependence on order-one factors such as 
sines and cosines of $\xi$ or $\et$ 
only changes sensitivity by factors of order unity, 
and is therefore relatively unimportant. 

\begin{table} 
\setlength{\tabcolsep}{0.5mm} 
\renewcommand{\arraystretch}{0.5} 
\begin{ruledtabular} 
\begin{tabular}{c|c|*8{c}} 
 & $\om_\oplus T_\oplus$ 
 & & & & & & & & 
\\
 & dep. 
 & $\tc_Q$ & $\tc_X$ & $\tc_Y$ & $\tc_Z$ & $\tc_-$ & $\tc_{TX}$ & $\tc_{TY}$ & $\tc_{TZ}$ 
\\ \hline 
% begin tilde c_q dependence =========================================== 
& & &&&&&&& \\ 
$\tc_{\bar q}-\tc_q$ & $\codt$  
 & \beo & \beo & \nop & \beo & \beo & \nop & \bet & \nop 
\\ 
 & $\sodt$ 
 & \beo & \nop & \beo & \beo & \beo & \bet & \nop & \nop 
\\ 
 & $\ctodt$ 
 & \nop & \nop & \nop & \nop & \bes  & \bto & \bto & \nop 
\\ 
 & $\stodt$ 
 & \nop & \nop & \nop & \bes  & \nop & \bto & \bto & \nop 
\\ 
 & const.  
 & \bes  & \nop & \nop & \nop & \nop & \bto & \bto & \bto 
\\ & & &&&&&&& 
\end{tabular} 
\end{ruledtabular} 
\caption{Dependance of $\tc_{\bar q}-\tc_q$ on the lab's rotational frequency $\om_\oplus$ as Earth rotates, 
 on Sun-frame tilde coefficients, 
 and on various speeds.} 
\label{ctable} 
\end{table} 

The relevant content is summarized in Table \ref{ctable}. 
Each column corresponds to a different Sun-frame tilde combination of coefficients for Lorentz violation, 
while each row corresponds to a type of time dependence. 
The entries in the table give the suppression factor for the dominant contribution 
of each Sun-frame tilde combination to each type of time dependence. 
For example, the 
$-\frac{1}{3}\be\be_\oplus\ce\cto\tc_Q\codt$ term in \Eq{Decq} 
is represented by the upper-left $\be\be_\oplus$ entry in Table \ref{ctable}. 
Note that the $\frac{1}{3}\be\be_L\tc_Q$ constant term is not represented in the table, 
as it is dominated by the $\frac{1}{6}\be^2\tc_Q$ term. 

Detailed expressions like \Eq{Decq} for other coefficients 
are unwieldy and give no particular insight. 
The relevant information is displayed two ways. 
First, the time and Sun-frame-component dependence is summarized in Tables \ref{bdggtable1} and \ref{bdggtable2}, 
which are entirely analogous to Table \ref{ctable}. 

Second, the detailed results in the limits $\be_L\rightarrow 0$, $\be_\oplus\rightarrow 0$ are given. 
This corresponds to neglecting relativistic effects 
associated with the lab's motion through space. 
It is equivalent to assuming that the lab has a constant velocity, 
but rotates about an axis parallel to Earth's. 
Since $\be_L\ll\be$ and $\be_\oplus\ll\be$ in any current experiment, 
all highest-order effects are represented. 
The coefficient differences listed in \Eq{DeltaTildes1}, 
when expressed in Sun-frame components, are given in the Appendix. 

\begin{table*}
\renewcommand{\arraystretch}{0.5} 
\tabletextsize 
\begin{ruledtabular} 
\begin{tabular}{c|c|*3{c}*3{c}*4{c}*4{c}*3{c} } 
 & $\om_\oplus T_\oplus$ dep. 
 & $\tb_X$ & $\tb_Y$ & $\tb_Z$ 
 & $\td_X$ & $\td_Y$ & $\td_Z$ 
 & $\tg_{TX}$ & $\tg_{TY}$ & $\tg_{TZ}$ & $\tg_{-}$ 
 & $\tg_{DX}$ & $\tg_{DY}$ & $\tg_{DZ}$ & $\tg_Q$ 
 & $\tH_{XT}$ & $\tH_{YT}$ & $\tH_{ZT}$ 
\\ \hline 
% tilde b_3 ======== 
&& &&& &&& &&&& &&&& && \\ 
$\tb_{\bar 3}-\tb_3$ & $\codt$  
%  \tb_X                 \td_X                 \tg_TX                       \tg_DX                       \tH_XT             
 & \bes & \beo & \beo  & \nop & \beo & \nop  & \bes & \beo & \beo & \beo  & \bes & \beo & \beo & \nop  & \bet & \bto & \bto \\  & $\sodt$ 
 & \beo & \bes & \beo  & \beo & \nop & \nop  & \beo & \bes & \beo & \beo  & \beo & \bes & \beo & \nop  & \bto & \bet & \bto \\  & $\ctodt$ 
 & \beo & \beo & \nop  & \beo & \beo & \nop  & \beo & \beo & \bes & \beo  & \beo & \beo & \nop & \nop  & \nop & \nop & \nop \\  & $\stodt$ 
 & \beo & \beo & \nop  & \beo & \beo & \nop  & \beo & \beo & \beo & \bes  & \beo & \beo & \nop & \nop  & \nop & \nop & \nop \\  & const.  
 & \beo & \beo & \bes  & \beo & \beo & \nop  & \beo & \beo & \nop & \nop  & \beo & \beo & \bes & \beo  & \bto & \bto & \bet \\ && &&& &&& &&&& &&&& && \\ \hline 
% tilde d_3 ========                                                                                  
&& &&& &&& &&&& &&&& && \\                                                                            
$\td_{\bar 3}-\td_3$ & $\codt$                                                                        
 & \nop & \beo & \nop  & \bes & \beo & \beo  & \nop & \nop & \nop & \nop  & \nop & \nop & \nop & \nop  & \bet & \bto & \bto \\  & $\sodt$ 
 & \beo & \nop & \nop  & \beo & \bes & \beo  & \nop & \nop & \nop & \nop  & \nop & \nop & \nop & \nop  & \bto & \bet & \bto \\  & $\ctodt$ 
 & \beo & \beo & \nop  & \beo & \beo & \nop  & \nop & \nop & \nop & \nop  & \nop & \nop & \nop & \nop  & \nop & \nop & \nop \\  & $\stodt$ 
 & \beo & \beo & \nop  & \beo & \beo & \nop  & \nop & \nop & \nop & \nop  & \nop & \nop & \nop & \nop  & \nop & \nop & \nop \\  & const.  
 & \beo & \beo & \nop  & \beo & \beo & \bes  & \nop & \nop & \nop & \nop  & \nop & \nop & \nop & \nop  & \bto & \bto & \bet \\ && &&& &&& &&&& &&&& && \\ \hline 
% tilde g_d ========                                                                                  
&& &&& &&& &&&& &&&& && \\                                                                            
$\tg_{\bar d}-\tg_d$ & $\codt$                                                                        
 & \bes & \beo & \beo  & \nop & \nop & \nop  & \bes & \beo & \beo & \beo  & \bes & \beo & \beo & \nop  & \nop & \nop & \nop \\  & $\sodt$ 
 & \beo & \bes & \beo  & \nop & \nop & \nop  & \beo & \bes & \beo & \beo  & \beo & \bes & \beo & \nop  & \nop & \nop & \nop \\  & $\ctodt$ 
 & \beo & \beo & \nop  & \nop & \nop & \nop  & \beo & \beo & \bes & \beo  & \beo & \beo & \nop & \nop  & \nop & \nop & \nop \\  & $\stodt$ 
 & \beo & \beo & \nop  & \nop & \nop & \nop  & \beo & \beo & \beo & \bes  & \beo & \beo & \nop & \nop  & \nop & \nop & \nop \\  & const.  
 & \beo & \beo & \bes  & \nop & \nop & \nop  & \beo & \beo & \nop & \nop  & \beo & \beo & \bes & \beo  & \nop & \nop & \nop \\ && &&& &&& &&&& &&&& && \\ \hline 
% tilde g_q=========                                                                                  
&& &&& &&& &&&& &&&& && \\                                                                            
$\tg_{\bar q}-\tg_q$ & $\codt$                                                                        
 & \beo & \nop & \beo  & \nop & \nop & \nop  & \beo & \bes & \beo & \beo  & \beo & \nop & \beo & \beo  & \nop & \nop & \nop \\  & $\sodt$ 
 & \nop & \beo & \beo  & \nop & \nop & \nop  & \bes & \beo & \beo & \beo  & \nop & \beo & \beo & \beo  & \nop & \nop & \nop \\  & $\ctodt$ 
 & \beo & \beo & \nop  & \nop & \nop & \nop  & \beo & \beo & \beo & \bes  & \beo & \beo & \nop & \nop  & \nop & \nop & \nop \\  & $\stodt$ 
 & \beo & \beo & \nop  & \nop & \nop & \nop  & \beo & \beo & \bes & \beo  & \beo & \beo & \nop & \nop  & \nop & \nop & \nop \\  & const.  
 & \beo & \beo & \beo  & \nop & \nop & \nop  & \beo & \beo & \nop & \nop  & \beo & \beo & \beo & \bes  & \nop & \nop & \nop \\ && &&& &&& &&&& &&&& && 
\end{tabular} 
\end{ruledtabular} 
\caption{Dependance of $\tb_{\bar 3}-\tb_3$, $\td_{\bar 3}-\td_3$, $\tg_{\bar d}-\tg_d$, $\tg_{\bar q}-\tg_q$
 on the lab's rotational frequency $\om_\oplus$ as Earth rotates, 
 on Sun-frame tilde coefficients, 
 and on various speeds 
 (continued in Table \ref{bdggtable2}).} 
\label{bdggtable1}
\end{table*} 

\begin{table*}
\renewcommand{\arraystretch}{0.5} 
\tabletextsize 
\begin{ruledtabular} 
\begin{tabular}{c|c|*3{c}*6{c}*6{c} } 
 & $\om_\oplus T_\oplus$ dep. 
 & $\td_{XY}$ & $\td_{YZ}$ & $\td_{ZX}$ 
 & $\td_Q$ & $\td_+$ & $\td_-$ & $\tb_T$ & $\tg_T$ & $\tg_c$ 
 & $\tg_{XY}$ & $\tg_{XZ}$ & $\tg_{YX}$ & $\tg_{YZ}$ & $\tg_{ZX}$ & $\tg_{ZY}$ 
\\ \hline 
% tilde b_3 ======== 
&&& &&&&&& &&&&& \\ 
$\tb_{\bar 3}-\tb_3$ & $\codt$  
%  \td_XY                \td_Q                                      \tg_XY                      
 & \bto & \nop & \nop  & \bto & \bto & \bto & \bto & \bto & \bto  & \btl & \nop & \nop & \bto & \nop & \bto \\  & $\sodt$ 
 & \nop & \bto & \bet  & \bto & \bto & \bto & \bto & \bto & \bto  & \nop & \bto & \btl & \nop & \bto & \nop \\  & $\ctodt$ 
 & \bet & \nop & \nop  & \nop & \nop & \nop & \bto & \bto & \bto  & \bto & \nop & \bto & \nop & \btl & \btl \\  & $\stodt$ 
 & \nop & \nop & \nop  & \nop & \nop & \bet & \bet & \bet & \bet  & \bto & \nop & \bto & \nop & \bto & \bto \\  & const.  
 & \bet & \nop & \bto  & \bto & \bto & \nop & \nop & \nop & \nop  & \bto & \nop & \bto & \nop & \btl & \btl \\ &&& &&&&&& &&&&& \\ \hline 
% tilde d_3 ========                                             
&&& &&&&&& &&&&& \\                                               
$\td_{\bar 3}-\td_3$ & $\codt$                                   
 & \bto & \bet & \bto  & \bto & \bto & \bto & \bto & \bto & \nop  & \bet & \bet & \bto & \bto & \bto & \bto \\  & $\sodt$ 
 & \bto & \bto & \bet  & \bto & \bto & \bto & \bto & \bto & \nop  & \bto & \bto & \bet & \bet & \bto & \bto \\  & $\ctodt$ 
 & \bet & \nop & \nop  & \nop & \nop & \nop & \nop & \nop & \nop  & \nop & \nop & \nop & \nop & \bet & \bet \\  & $\stodt$ 
 & \nop & \nop & \nop  & \nop & \nop & \bet & \nop & \nop & \nop  & \nop & \nop & \nop & \nop & \nop & \nop \\  & const.  
 & \bet & \bto & \bto  & \bto & \bto & \nop & \bto & \bto & \nop  & \bto & \bto & \bto & \bto & \bet & \bet \\ &&& &&&&&& &&&&& \\ \hline 
% tilde g_d ========                                             
&&& &&&&&& &&&&& \\                                               
$\tg_{\bar d}-\tg_d$ & $\codt$                                   
 & \nop & \nop & \nop  & \nop & \nop & \nop & \bto & \bto & \bto  & \bet & \nop & \nop & \bto & \nop & \bto \\  & $\sodt$ 
 & \nop & \nop & \nop  & \nop & \nop & \nop & \bto & \bto & \bto  & \nop & \bto & \bet & \nop & \bto & \nop \\  & $\ctodt$ 
 & \nop & \nop & \nop  & \nop & \nop & \nop & \bto & \bto & \bto  & \bto & \nop & \bto & \nop & \bet & \bet \\  & $\stodt$ 
 & \nop & \nop & \nop  & \nop & \nop & \nop & \bet & \bet & \bet  & \bto & \nop & \bto & \nop & \bto & \bto \\  & const.  
 & \nop & \nop & \nop  & \nop & \nop & \nop & \bto & \bto & \nop  & \bto & \nop & \bto & \nop & \bet & \bet \\ &&& &&&&&& &&&&& \\ \hline 
% tilde g_q=========                                             
&&& &&&&&& &&&&& \\                                               
$\tg_{\bar q}-\tg_q$ & $\codt$                                   
 & \nop & \nop & \nop  & \nop & \nop & \nop & \bto & \bto & \bto  & \nop & \bto & \bet & \bet & \bto & \nop \\  & $\sodt$ 
 & \nop & \nop & \nop  & \nop & \nop & \nop & \nop & \nop & \bto  & \bet & \bet & \nop & \bto & \nop & \bto \\  & $\ctodt$ 
 & \nop & \nop & \nop  & \nop & \nop & \nop & \bet & \bet & \bet  & \bto & \nop & \bto & \nop & \bto & \bto \\  & $\stodt$ 
 & \nop & \nop & \nop  & \nop & \nop & \nop & \bto & \bto & \bto  & \bto & \nop & \bto & \nop & \bet & \bet \\  & const.  
 & \nop & \nop & \nop  & \nop & \nop & \nop & \bet & \bet & \nop  & \bto & \bto & \bto & \bto & \bto & \bto \\ &&& &&&&&& &&&&& 
\end{tabular} 
\end{ruledtabular} 

\caption{Dependance of $\tb_{\bar 3}-\tb_3$, $\td_{\bar 3}-\td_3$, $\tg_{\bar d}-\tg_d$, $\tg_{\bar q}-\tg_q$
 on the lab's rotational frequency $\om_\oplus$ as Earth rotates, 
 on Sun-frame tilde coefficients, 
 and on various speeds 
 (continued from Table \ref{bdggtable1}).} 
\label{bdggtable2}
\end{table*} 

%======================================================================================
\section{The Heidelberg Heavy-Ion Experiment} 
\citelabel{Heidelberg} 
%======================================================================================
The recent experiment at the Heidelberg heavy-ion storage facility \cite{doppler1} 
used saturation spectroscopy within $^7$Li$^+$ ions moving at speed $\be\approx 0.064$ 
to probe the Doppler effect. 
The experiment effectively averaged over all of the $\De m_F=0$ transitions 
between the $^3P_2 (F=7/2)$ and $^3S_1 (F=5/2)$ levels, 
to each of which the analysis of Sec.\ \ref{SMEPredictions} may be applied. 
Thus, the measured frequency shift in each group of atoms is 
\bea 
\left< \de\nu \right> 
 &=& \frac{1}{6} \sum_{m_F=-5/2}^{+5/2} \left[ 
 \fr{\de E(F=\frac{7}{2},m_F)-\de E(F=\frac{5}{2},m_F)}{2\pi} 
 \right] \nonumber \\ 
&=& \frac{1}{12\pi} \sum_{m_F=-5/2}^{+5/2} \Big[ 
 \hmf E_d(F=\frac{7}{2})+\tmf E_q(F=\frac{7}{2}) \nonumber \\ 
 && \quad\quad 
 -\pr{\hmf} E_d(F=\frac{5}{2})-\pr{\tmf} E_q(F=\frac{5}{2})
 \Big]  
\quad . 
\labeleea{AvgDenu1} 
Since $\hmf$ is linear in $m_F$, all $E_d$ contributions to this sum cancel each other out. 
Using the fact that any sum of $\hmf$ or $\tmf$ over a complete set 
of $m_F=-F, \cdots, +F$ is zero \cite{SMEclocks1}, 
the $F=\frac{5}{2}$ parts of this sum contribute nothing. 
Finally, the $E_q(F=\frac{7}{2})$ sum can be explicitly tallied to give 
\beq 
\left< \de\nu \right> = -\frac{1}{6\pi} E_q(F=\frac{7}{2}) 
 = -\frac{1}{6\pi} \sum_w 
 \left( 
 \ga_w\tc_q^w+\la_w\tg_q^w 
 \right) 
\quad . 
\labeleeq{AvgDenu2} 
The key experimental quantity for the Heidelberg experiment is then 
\beq 
\fr{\bar\nu_{\rm beam}^2}{\nu_{\rm lab}^2} = 1 
 -\fr{1}{3\pi\nu_{\rm SM}} \sum_{w} 
 \left[ 
 \ga_w (\tc_{\bar q}^w-\tc_q^w) 
 +\la_w (\tg_{\bar q}^w-\tg_q^w) 
 \right] 
\quad . 
\labeleeq{Heiderberg1} 

Though calculation of the $\ga_w$ and $\la_w$ parameters for nucleons 
may in general be quite involved \cite{SMEclocks1}, 
the nuclear Schmidt model suffices for this work. 
For $^7$Li$^+$ in the $F=\frac{7}{2}$ state, 
the Schmidt model gives $\ga_p\approx -\frac{1}{15}\times 10^{-2}$ and $\ga_n=0$, 
while standard atomic calculations lead to $\ga_e\approx -\frac{1}{15}\times 10^{-5}$. 
For each particle species, Schmidt model and standard atomic calculations give $\la_w=0$, 
so $\tg_{\bar q}-\tg_q$ makes no contribution. 

An expression for $\tc_{\bar q}-\tc_q$ 
in terms of nonrotating-frame components is given by \Eq{Decq}. 
%While detailed timing data was not kept in the Heidelberg experiment, 
%frequency measurements were effectively averaged over several days \cite{gwinner}. 
The Heidelberg experiment effectively averaged frequency measurements over several days \cite{gwinner}, 
so the contributions proportional to sines and cosines involving sidereal frequency $\om_\oplus$ 
all average to approximately zero \cite{FootnoteSiderealSines}. 
Hence, the effective value of $\tc_{\bar{q}}-\tc_q$ is 
\bea 
\tc_{\bar{q}}-\tc_{q} 
 &=& \be^2\be_\oplus \left( 
 \frac{3}{2}\sto\tc_{TX}-\frac{3}{2}\ce\cto\tc_{TY}-\se\cto\tc_{TZ}\right) 
  \nonumber \\ 
&& +\be^2 \left(\frac{1}{6}\tc_Q\right) + \be\be_L \left(\frac{1}{3}\tc_Q\right) 
\quad . 
\labeleea{Heidelberg3} 
Note that this result could be derived from Table \ref{ctable}, 
though several order-one factors would be missing. 
Also missing would be the last term. 
However, since it is dominated by the next-to-last term, 
it is irrelevant. 
It is neglected in the following calculations. 
 
The key quantity for the Heidelberg experiment is then 
\bea 
\fr{\bar\nu_{\rm beam}^2}{\nu_{\rm lab}^2} 
 &=& 1- \fr{1}{3\pi\nu_{\rm SM}} \sum_{w=p,e} \ga_w \Big[ \frac{1}{6}\be^2\tc_Q^w 
 + \be^2\be_\oplus \big( 
 \frac{3}{2}\sto\tc_{TX}^w \nonumber \\ 
&& \qquad\qquad\qquad -\frac{3}{2}\ce\cto\tc_{TY}^w-\se\cto\tc_{TZ}^w\big) \Big]
\quad . 
\labeleea{Heidelberg4} 
Due to experimental data being taken over a small number of days, 
$\cto$ and $\sto$ are each approximately constants of order one, 
as are $\ce$ and $\se$. 
Comparison with \Eq{bound1} 
and expression of the coefficient combinations in terms of original SME coefficients 
(i.e., in terms of ``nontilde'' coefficients) 
leads to rough bounds 
\bea 
\abs{c_{XX}^p+c_{YY}^p-2c_{ZZ}^p} &\lsim& 10^{-11} \quad , \nonumber \\ 
\abs{c_{TJ}^p+c_{JT}^p} &\lsim& 10^{-8} \quad (J=X,Y,Z) \quad , \nonumber \\ 
\abs{c_{XX}^e+c_{YY}^e-2c_{ZZ}^e} &\lsim& 10^{-5} \quad , \quad \mbox{and} \nonumber \\ 
\abs{c_{TJ}^e+c_{JT}^e} &\lsim& 10^{-2} \quad (J=X,Y,Z) 
\quad . 
\labeleea{bound2} 
Though detailed calculations have not been performed, 
more realistic nuclear models likely give similar-size bounds to many other parameters, 
including $\tc_Q$ for the neutron and 
$(\tg_{\bar q}-\tg_q)$-associated coefficients for both the proton and neutron. 

Nearly all of these bounds are new 
--- only $c_{XX}+c_{YY}-2c_{ZZ}$ for the electron has been previously bounded. 
They complement recent bounds on certain electron $c_{MN}$ combinations \cite{SMEelectrons2}, 
spin-torsion and Penning-trap bounds on other electron coefficients \cite{EXPTelectrons}, 
and dipole-type bounds on proton and electron coefficients derived from clock-comparison experiments \cite{SMEclocks1,EXPTclocks}. 

This work neglects effects from the photon sector of the SME. 
However, such effects \cite{tobar} can be readily appended to this analysis. 
Under the assumptions and notation of this work, 
photon contributions to the key experimental quantity would appear 
as an additional term 
$4\be\be_{\oplus}\se\sx\cto\tilde{\kappa}_{\rm tr}$ in \Eq{Heidelberg4}. 
This leads to the bound on $\tilde{\kappa}_{\rm tr}$ quoted in the earlier work.

%======================================================================================
\section{Future Experiments} 
\citelabel{FutureExperiments} 
There are two broad classes of changes that could be made to future Doppler-effect experiments 
to increase the set of coefficient sensitivity 
while using the same atomic species. 
The first involves study of a different set of transitions, 
while the second involves the time dependence of frequency measurements. 

First, an experiment studying a different set of transitions 
could be sensitive to several dipole-type coefficient differences 
that appear in \Eq{exptquant2}. 
The different transition set could be a reduced subset of $\De m_F=0$ transitions, 
or a set of $\De m_F\ne 0$ transitions. 
Calculations like those that appear in Sec.\ \ref{Heidelberg} of the current work 
may be used to determine the set of coefficient differences 
to which a given experiment is sensitive. 
For example, the standard clock transition $m_F=0 \rightarrow \pr{m_F}=0$ 
is incapable of measuring any Lorentz-violation effects. 

Changes of the first type 
would induce the relevance of more rows of Tables \ref{bdggtable1} and \ref{bdggtable2}. 
Without incorporating the time-dependence change described below, 
only those associated with time-constant contributions would appear. 
This could lead to sensitivity to all Sun-frame combinations except 
$\tg_{TZ}$, $\tg_{-}$, $\tg_c$, and $\td_{-}$ from Tables \ref{bdggtable1} and \ref{bdggtable2} 
and $\tc_X$, $\tc_Y$, $\tc_Z$, and $\tc_-$ from Table \ref{ctable}. 
Sensitivity to these coefficient combinations requires implementation of the second change. 

Second, timing information could be kept to allow study of the key experimental quantity 
as the lab rotates and accelerates through the Sun frame. 
This would lead to sensitivity to the sinusoidal rows of Table \ref{ctable}. 
Combination of this change together with the first 
could lead to sensitivity to the sinusoidal rows of Tables \ref{bdggtable1} and \ref{bdggtable2} as well. 

%======================================================================================

%======================================================================================
\section{Summary}
\citelabel{Summary} 
%======================================================================================
%This paper consists of wide speculations, vague definitions, implausible approximations, 
%and wretched grammar.  We apologize for wasting the reader's time. 
This work describes an analysis of Doppler-shift experiments in the context of Lorentz violation, 
conducted within the framework of the Standard-Model Extension. 
Included are effects up to second order in the beam atoms' speed with respect to the lab, 
first order in Earth's speed with respect to the Sun. 
and first order in the lab's speed with respect to Earth's rotational axis. 
The key experimental quantity is expressed in terms of 
combinations of nonrotating-Sun-frame Lorentz-violation coefficients, 
atomic/nuclear expectation values, geometrical factors, and time. 

The most recent Doppler-shift experiment at the Heidelberg storage ring 
is studied in detail. 
It is found to yield constraints on several previously unprobed 
Lorentz-violation coefficients associated with protons 
at the level of $10^{-11}$ and $10^{-8}$, 
and coefficients associated with electrons 
at the level of $10^{-2}$. 
Bounds on many other coefficients for neutrons and protons 
would likely result from more detailed nuclear calculations. 

%=================================================================
%=================================================================

\begin{acknowledgments} 
C.D.L.\ thanks Gerald Gwinner and Alan Kosteleck\'y for useful discussions. 
\end{acknowledgments}

%======================================================================================
\appendix* 
\section{Lab-frame coefficient differences} 
\label{CoeffDiff} 
%======================================================================================

Below are given 
the coefficient differences listed in \Eq{DeltaTildes1} 
in the limits $\be_\oplus\rightarrow0$, $\be_L\rightarrow 0$, 
when expressed in Sun-frame components. 
\begin{widetext} 
\bea 
%=====================================================================================
% Begin \tb_3 
%=====================================================================================
\tb_{\bar 3}-\tb_3 &=& 
% Begin cos(\om T_s) dependence====================================
\codt \Big\{ \be\Big[
 -\cx\tH_{XT}  
 \Big]
 +\be^2 \Big[ 
 \half\sx\bx +\half\sx\gdx -\half\sx\gtx 
 \Big] 
% \nonumber \\ 
%&& +\be\be_\oplus \Big[ 
% \cx\sto\tg_- +\half\cx\cto\se\tb_Y -\ce\cx\cto\tb_Z +\cx\cto\se\td_Y +\half\cx\cto\se\tg_{DY} 
% -\ce\cx\cto\tg_{DZ} -\half\cx\cto\se\tg_{TY} \nonumber \\ 
% && \qquad -\ce\cx\cto\tg_{TZ}
% \Big] 
% \nonumber \\ 
%&& +\be^2\be_\oplus \Big[ 
% -\frac{1}{4}\sx\sto\dm -\sx\sto\dpl -\frac{5}{4}\sx\sto\bt +\frac{1}{4}\sx\sto\dq 
% +\half\ce\cto\sx\dxy +2\sx\sto\gc +\frac{5}{4}\sx\sto\gt \nonumber \\ 
% && \qquad -\half\cto\se\sx\gyz 
% +\frac{3}{2}\ce\cto\sx\gzy +\half\cto\se\sx\hyt -\half\ce\cto\sx\hzt 
% \Big] 
 \Big\} 
 \nonumber \\ 
% Begin sin(\om T_s) dependence====================================
&+& \sodt \Big\{ \be\Big[ 
 \cx\dzx -\cx\hyt
 \Big]
 +\be^2 \Big[ 
 \half\sx\by +\half\sx\gdy +\half\sx\gty 
 \Big] 
% \nonumber \\ 
%&& +\be\be_\oplus \Big[ 
% \Big] 
% \nonumber \\ 
%&& +\be^2\be_\oplus \Big[ 
% \Big] 
 \Big\} 
 \nonumber \\ 
% Begin cos(2\om T_s) dependence====================================
&+& \ctodt \Big\{ \be\Big[
 -\half\sx\dxy 
 \Big]
 +\be^2 \Big[ 
 \half\cx\gtz 
 \Big] 
% \nonumber \\ 
%&& +\be\be_\oplus \Big[ 
% \Big] 
% \nonumber \\ 
%&& +\be^2\be_\oplus \Big[ 
% \Big] 
 \Big\} 
 \nonumber \\ 
% Begin sin(2\om T_s) dependence====================================
&+& \stodt \Big\{ \be\Big[
 \half\sx\dm +\half\sx\bt -\sx\gc -\half\sx\gt 
 \Big]
 +\be^2 \Big[ 
 -\half\cx\gm 
 \Big] 
% \nonumber \\ 
%&& +\be\be_\oplus \Big[ 
% \Big] 
% \nonumber \\ 
%&& +\be^2\be_\oplus \Big[ 
% \Big] 
 \Big\} 
 \nonumber \\ 
% Begin T independence====================================
&+& \Big\{ \be\Big[ 
 -\half\sx\dxy +\sx\hzt 
 \Big]
 +\be^2 \Big[ 
 \half\cx\bz +\half\cx\gdz 
 \Big] 
% \nonumber \\ 
%&& +\be\be_\oplus \Big[ 
% \Big] 
% \nonumber \\ 
%&& +\be^2\be_\oplus \Big[ 
% \Big] 
 \Big\} \quad , 
\labeleea{Explicitb3}  
\bea 
%=====================================================================================
% Begin \td_3 
%=====================================================================================
\td_{\bar 3}-\td_3 &=& 
% Begin cos(\om T_s) dependence====================================
\codt \Big\{ \be\Big[
 \cx\dyz+\half\cx\gxy-\cx\gxz-\half\cx\hxt 
 \Big]
 +\be^2 \Big[ 
 \half\sx\dx 
 \Big] 
% \nonumber \\ 
%&& +\be\be_\oplus \Big[ 
% \Big] 
% \nonumber \\ 
%&& +\be^2\be_\oplus \Big[ 
% \Big] 
 \Big\} 
 \nonumber \\ 
% Begin sin(\om T_s) dependence====================================
&+& \sodt \Big\{ \be\Big[ 
 -\half\cx\dzx +\half\cx\gyx -\cx\gyz -\half\cx\hyt 
 \Big]
 +\be^2 \Big[ 
 \half\sx\dy 
 \Big] 
% \nonumber \\ 
%&& +\be\be_\oplus \Big[ 
% \Big] 
% \nonumber \\ 
%&& +\be^2\be_\oplus \Big[ 
% \Big] 
 \Big\} 
 \nonumber \\ 
% Begin cos(2\om T_s) dependence====================================
&+& \ctodt \Big\{ \be\Big[ 
 \frac{3}{4}\sx\dxy +\frac{3}{4}\sx\gzx-\frac{3}{4}\sx\gzy 
 \Big]
% \nonumber \\ 
%&& +\be\be_\oplus \Big[ 
% \Big] 
% \nonumber \\ 
%&& +\be^2\be_\oplus \Big[ 
% \Big] 
 \Big\} 
 \nonumber \\ 
% Begin sin(2\om T_s) dependence====================================
&+& \stodt \Big\{ \be\Big[ 
 -\frac{3}{4}\sx\dm 
 \Big]
% \nonumber \\ 
%&& +\be\be_\oplus \Big[ 
% \Big] 
% \nonumber \\ 
%&& +\be^2\be_\oplus \Big[ 
% \Big] 
 \Big\} 
 \nonumber \\ 
% Begin T independence====================================
&+& \Big\{ \be\Big[ 
 -\frac{1}{4}\sx\dxy+\frac{1}{4}\sx\gzx +\frac{1}{4}\sx\gzy +\half\sx\hzt 
 \Big]
 +\be^2 \Big[ \half\cx\dz 
 \Big] 
% \nonumber \\ 
%&& +\be\be_\oplus \Big[ 
% \Big] 
% \nonumber \\ 
%&& +\be^2\be_\oplus \Big[ 
% \Big] 
 \Big\} \quad , 
\labeleea{Explicitd3}  
\bea 
%=====================================================================================
% Begin \tg_d 
%=====================================================================================
\tg_{\bar d}-\tg_d &=& 
% Begin cos(\om T_s) dependence====================================
\codt \Big\{ \be\Big[ 
 2\cx\gxy 
 \Big]
 +\be^2 \Big[ 
 \half\sx\bx +\sx\gdx -\sx\gtx 
 \Big] 
% \nonumber \\ 
%&& +\be\be_\oplus \Big[ 
% \Big] 
% \nonumber \\ 
%&& +\be^2\be_\oplus \Big[ 
% \Big] 
 \Big\} 
 \nonumber \\ 
% Begin sin(\om T_s) dependence====================================
&+& \sodt \Big\{ \be\Big[ 
 2\cx\gyx 
 \Big]
 +\be^2 \Big[ 
 \half\sx\by +\sx\gdy +\sx\gty 
 \Big] 
% \nonumber \\ 
%&& +\be\be_\oplus \Big[ 
% \Big] 
% \nonumber \\ 
%&& +\be^2\be_\oplus \Big[ 
% \Big] 
 \Big\} 
 \nonumber \\ 
% Begin cos(2\om T_s) dependence====================================
&+& \ctodt \Big\{ \be\Big[ 
 \sx\gzx -\sx\gzy 
 \Big]
 +\be^2 \Big[ 
 \cx\gtz 
 \Big] 
% \nonumber \\ 
%&& +\be\be_\oplus \Big[ 
% \Big] 
% \nonumber \\ 
%&& +\be^2\be_\oplus \Big[ 
% \Big] 
 \Big\} 
 \nonumber \\ 
% Begin sin(2\om T_s) dependence====================================
&+& \stodt \Big\{ \be\Big[ 
 \sx\bt -2\sx\gc -\sx\gt 
 \Big]
 +\be^2 \Big[ -\cx\gm 
 \Big] 
% \nonumber \\ 
%&& +\be\be_\oplus \Big[ 
% \Big] 
% \nonumber \\ 
%&& +\be^2\be_\oplus \Big[ 
% \Big] 
 \Big\} 
 \nonumber \\ 
% Begin T independence====================================
&+& \Big\{ \be\Big[ 
 -\sx\gzx -\sx\gzy 
 \Big]
 +\be^2 \Big[ 
 \half\cx\bz +\cx\gdz 
 \Big] 
% \nonumber \\ 
%&& +\be\be_\oplus \Big[ 
% \Big] 
% \nonumber \\ 
%&& +\be^2\be_\oplus \Big[ 
% \Big] 
 \Big\} \quad , 
\labeleea{Explicitgd}  
\bea 
%=====================================================================================
% Begin \tc_q 
%=====================================================================================
\tc_{\bar q}-\tc_q &=& 
% Begin cos(\om T_s) dependence====================================
\codt \Big\{ \be\Big[ 
 \tcty 
 \Big]
 \Big\} 
% Begin sin(\om T_s) dependence====================================
+ \sodt \Big\{ \be\Big[ 
 -\tctx 
 \Big]
 \Big\} 
 \nonumber \\ 
% Begin cos(2\om T_s) dependence====================================
&+& \ctodt \Big\{ 
 \be^2 \Big[ 
 -\half\tcm 
 \Big] 
 \Big\} 
% Begin sin(2\om T_s) dependence====================================
+ \stodt \Big\{ 
 \be^2 \Big[ 
 -\half\tcz 
 \Big] 
 \Big\} 
% Begin T independence====================================
+ \Big\{ 
 \be^2 \Big[ 
 \frac{1}{6}\tcq 
 \Big] 
 \Big\} \quad , \quad {\rm and} 
\labeleea{Explicitcq}  
\bea 
%=====================================================================================
% Begin \tg_q 
%=====================================================================================
\tg_{\bar q}-\tg_q &=& 
% Begin cos(\om T_s) dependence====================================
\codt \Big\{ \be\Big[ 
 \big(\half-\frac{3}{2}\ctx\big)\gyx 
 +\big(\half+\frac{3}{2}\ctx\big)\gyz 
 \Big]
 +\be^2 \Big[ 
 -\frac{3}{4}\stx\gty 
 \Big] 
% \nonumber \\ 
%&& +\be\be_\oplus \Big[ 
% \Big] 
% \nonumber \\ 
%&& +\be^2\be_\oplus \Big[ 
% \Big] 
 \Big\} 
 \nonumber \\ 
% Begin sin(\om T_s) dependence====================================
&+& \sodt \Big\{ \be\Big[ 
 \big(-\half+\frac{3}{2}\ctx\big)\gxy 
 +\big(-\half-\frac{3}{2}\ctx\big)\gxz 
 \Big]
 +\be^2 \Big[ 
 \frac{3}{4}\stx\gtx 
 \Big] 
% \nonumber \\ 
%&& +\be\be_\oplus \Big[ 
% \Big] 
% \nonumber \\ 
%&& +\be^2\be_\oplus \Big[ 
% \Big] 
 \Big\} 
 \nonumber \\ 
% Begin cos(2\om T_s) dependence====================================
&+& \ctodt \Big\{ \be\Big[ 
 \frac{3}{4}\stx\bt +\frac{3}{2}\stx\gc +\frac{3}{4}\stx\gt 
 \Big]
 +\be^2 \Big[ 
 \big(-\frac{3}{8}+\frac{3}{8}\ctx\big)\gm 
 \Big] 
% \nonumber \\ 
%&& +\be\be_\oplus \Big[ 
% \Big] 
% \nonumber \\ 
%&& +\be^2\be_\oplus \Big[ 
% \Big] 
 \Big\} 
 \nonumber \\ 
% Begin sin(2\om T_s) dependence====================================
&+& \stodt \Big\{ \be\Big[ 
 \frac{3}{4}\stx\gzx -\frac{3}{4}\stx\gzy 
 \Big]
 +\be^2 \Big[ 
 \big(-\frac{3}{8}+\frac{3}{8}\ctx\big)\gtz 
 \Big] 
% \nonumber \\ 
%&& +\be\be_\oplus \Big[ 
% \Big] 
% \nonumber \\ 
%&& +\be^2\be_\oplus \Big[ 
% \Big] 
 \Big\} 
 \nonumber \\ 
% Begin T independence====================================
&+& \Big\{ \be\Big[ 
 -\frac{3}{4}\stx\bt +\frac{3}{4}\stx\gt 
 \Big]
 +\be^2 \Big[ 
 \big(\frac{1}{8}+\frac{3}{8}\ctx\big)\gq 
 \Big] 
% \nonumber \\ 
%&& +\be\be_\oplus \Big[ 
% \Big] 
% \nonumber \\ 
%&& +\be^2\be_\oplus \Big[ 
% \Big] 
 \Big\} . 
\labeleea{Explicitgq}  
\end{widetext}

%======================================================================================

\end{document}